\documentclass[11pt]{article}

\usepackage[latin1]{inputenc}
\usepackage[OT1]{fontenc}
\usepackage[english]{babel}
\usepackage{indentfirst}
\usepackage{aeguill}
\usepackage{epic,eepic}
\usepackage{graphicx}
\usepackage{epsfig}
\usepackage{amsthm, amsmath, amssymb}
\usepackage{setspace}
\usepackage[dvipsnames]{xcolor}
\usepackage{array}

\newtheorem{theorem}{Theorem}
\newtheorem{lemma}{Lemma}
\newtheorem{proposition}{Proposition}
\newtheorem{definition}{Definition}

\begin{document}

\bibliographystyle{alpha}

\title{Nonlinear threshold Boolean automata networks and phase transitions}

\author{Jacques Demongeot$^{1,}$\thanks{Jacques.Demongeot@imag.fr} \and Sylvain
  Sené$^{2,3,}$\thanks{Sylvain.Sene@ibisc.univ-evry.fr}}

\date{}
\maketitle

\begin{center}
  \vspace*{-9mm}{\small $^1~$Universit{\'e} Joseph Fourier de Grenoble,
    TIMC-IMAG, AGIM,
    Facult{\'e} de m{\'e}decine, 38700 La Tronche, France\\
    $^2~$Universit{\'e} d'{\'E}vry -- Val d'Essonne, IBISC, 91000
    {\'E}vry, France\\
    $^3~$Institut rh{\^o}ne-alpin des syst{\`e}mes complexes, IXXI, 69007 Lyon,
    France}\\[10mm]
\end{center}

\begin{abstract}
  In this report, we present a formal approach that addresses the problem of
  emergence of phase transitions in stochastic and attractive nonlinear
  threshold Boolean automata networks. Nonlinear networks considered are
  informally defined on the basis of classical stochastic threshold Boolean
  automata networks in which specific interaction potentials of neighbourhood
  coalition are taken into account. More precisely, specific nonlinear terms
  compose local transition functions that define locally the dynamics of such
  networks. Basing our study on nonlinear networks, we exhibit new results,
  from which we derive conditions of phase transitions.
\end{abstract}
 
\pagestyle{plain}
\pagenumbering{arabic}

\section{Introduction}
\label{intro}

The model of deterministic Threshold Boolean automata networks (called TBANs for
short in the sequel) has been developped in the 1940's by McCulloch and Pitts
in~\cite{McCulloch1943} as a way to represent logically the interactions between
neurons over time. In parallel~\cite{Onsager1944} has been addressed the problem
of existence of phase transition in the two-dimensional Ising model of
ferromagnetism~\cite{Ising1925}. Taking into account that the classical Ising
model can be generalised in the Boolean framework by the Boltzmann
machine~\cite{Ackley1985}, that is a stochastic variation around deterministic
TBANs, we propose in this report a partial solution of the problem of emergence
of phase transitions in this context, as it has been performed in the case of
the classical Ising model by Dobrushin and Ruelle
in~\cite{Dobrushin1968c,Ruelle1969}. More precisely, we present a generalisation
to nonlinear TBANs of theoretical results of phase transitions due to the
influence of fixed boundary conditions already obtained in the framework of
linear TBANs~\cite{djs08,ds08}.\medskip

After a presentation of important definitions for the study in
Section~\ref{def}, new theoretical results of phase transitions are given.

\section{Model definitions}
\label{def}

Although this work focuses on nonlinear TBANs whose architecture is partially
defined in a part of the lattice on $\mathbb{Z}^2$, let us present TBANs from
the general point of view. Let $N$ be such an arbitrary network. $N$ is composed
by $n$ nodes interacting over time through a labelled digraph $G = (V,A)$, where
$V$ is the set of nodes, elements of $\mathbb{Z}^2$, whose states are valued in
$\{0,1\}$ ($0$ when the node is inactive and $1$ when it is active) and $A
\subset V\times V$ is the set of arcs linking elements with each others. A TBAN
is characterised by:
\begin{itemize}
\item an \emph{interaction matrix} $W$ of order $n$: it defines the structure of
  $N$ and each coefficient $w_{i,j} \in \mathbb{R}$ is the label of arc $(j,i)$
  of $A$ and gives the interaction weight node $j$ has on node $i$. If $w_{i,j}$
  is null, then $(j,i) \notin A$, else node $j$ is said to be a neighbour of
  node $i$ and we note $j \in {\cal N}_i$. In this case, node $j$ is called an
  inducer/activator (resp. repressor/inhibitor) of node $i$ if $w_{i,j} > 0$
  (resp. $w_{i,j} < 0$);
\item a \emph{threshold vector} $\Theta$ of dimension $n$: each element
  $\theta_i$ is called the activation threshold of node $i$.
\item $n$ \emph{local transition functions} which define the local evolution of
  each of the nodes in the TBANs. The general concept of the local evolution of
  a node $i$, namely the calculation of its state at time $t+1$ being given $N$
  and the state of any node $k \in V$ at time $t$, is the following: if the
  potential of $i$ at time $t$, \textit{i.e.}, the sum of the interaction
  weights received from its active neighbours, is greater than (resp. not
  greater than) its activation threshold then its state at time $t+1$ equals $1$
  (resp. $0$). Thus, if we denote by $x_i(t)$ the state of node $i$ at time $t$,
  the local transitions functions are:
  \begin{equation}
    \label{eq_NN1}
    x_i(t+1) = {\cal H}(\sum_{j \in {\cal N}_i} w_{i,j} \cdot x_j(t) - \theta_i) 
    \text{,}
  \end{equation}
  where ${\cal H}$ represents the Heaviside (or sign-step) function and is such
  that ${\cal H}(x) = \begin{cases} 0 & \text{if } x \leq 0\text{,}\\ 1 &
    \text{otherwise.} \end{cases}$
\end{itemize}

An application $x: V \to \{0,1\}$ is called a \emph{configuration} of $N$. In
other words, the vector $x(t) = (x_i(t))_{i \in V} \in \{0,1\}^n$ is the
configuration of $N$ at time $t$.\medskip

In the sequel, in order to highlight the emergence of phase transitions from the
dynamical behaviour of TBANs, we will give a particular attention to the notion
of boundary conditions. We will explain this later. Nevertheless, since we focus
on TBANs on $\mathbb{Z}^2$, let us present general definitions of the notions of
center and boundary of a graph $G=(V,A)$ that we will be able to adapt in the
context of two-dimensional lattices. Basic notions of graph theory are
considered to be known (\textit{cf}.~\cite{Harary1969}).

\begin{definition}
  \label{def_boundary}
  Let $G=(V,A)$ an arbitrary digraph. The \emph{boundary} of $G$ is the set of
  its sources.
\end{definition}

Let $u$ and $v$ be two distinct vertices of a digraph $G = (V,A)$. The
\emph{distance} $d(u, v)$ is the length of the shortest path linking $u$ to
$v$. If there is no path from $u$ to $v$, $d(u,v)$ is defined as equal to
$+\infty$. 

\begin{definition}
  \label{def_eccentricity}
  Let $G=(V,A)$ an arbitrary digraph. The \emph{eccentricity} $\varepsilon(u)$
  of a non isolated vertex $u \in V$ is the maximal distance less than $+\infty$
  from $u$ and every other vertex of $G$, such that $\varepsilon(u) =
  \text{Max}_{v \, \in \, S \setminus u}(d(u,v) < +\infty)$.
\end{definition}

\begin{definition}
  \label{def_centre}
  Let $G=(V,A)$ an arbitrary digraph. The \emph{centre} of $G$ is the set of its
  vertices of minimal eccentricity.
\end{definition}

In this report, we differentiate the notions of neighbourhood and strict
neighbourhood of nonlinear two-dimentional TBANs according to the following
definitions.

\begin{definition}
  \label{def_neighbourhood}
  Let $N$ be a two-dimensional TBAN on $\mathbb{Z}^2$. The \emph{neighbourhood}
  ${\cal N}_i$ of node $i$ is the set composed of nearest-neighbours nodes
  (\textit{i.e.}, nodes at distance $1$ to $i$) of $i$ and $i$ itself.
\end{definition}

\begin{definition}
  \label{def_strictneighbourhood}
  Let $N$ be a two-dimensional TBAN on $\mathbb{Z}^2$. The \emph{strict
    neighbourhood} $\Lambda_i$ of node $i$ is such that $\Lambda_i = {\cal N}_i
  \setminus \{i\}$.
\end{definition}

Let us now define the properties of isotropy and translation invariance of the
two-dimensional TBANs considered.

\begin{definition}
  \label{def_isotropy}
  Let $N$ be a two-dimensional TBAN on $\mathbb{Z}^2$. $N$ is \emph{isotropic}
  if and only if: 
  \begin{equation*}
    \forall i \in N, \, \forall j, j' \in {\cal N}_i, \, w_{i,j} = w_{i,j'} 
    \text{.}
  \end{equation*}
\end{definition}

\begin{definition}
  \label{def_translationinvariance}
  Let $N$ be a two-dimensional TBAN on $\mathbb{Z}^2$. $N$ is \emph{translation
    invariant} if and only if, given $j_1, \ldots, j_k \in {\cal N}_i$, it holds
  that: 
  \begin{equation*}
    \forall i, i' \in N, \, \exists s \in \mathbb{Z}^d, \, i' = i+s, \,
    \forall \ell \in \{1,\ldots,k\}, \, j'_{\ell} = j_{\ell}+s:\: w_{i,j_{\ell}} =
    w_{i',j'_{\ell}}\text{.}
  \end{equation*}
\end{definition}

As a consequence, TBANs considered in this study are \emph{symmetric},
\textit{i.e.}, they are such that $\forall i, \forall j \in {\cal N}_i, w_{i,j}
= w_{j,i}$. According to these properties of isotropy and translation
invariance, it is easy to see that Definition~\ref{def_centre} can be applied
directly to nonlinear TBANs on $\mathbb{Z}^2$. Conversely, the set of boundary
obtained from the application of Definition~\ref{def_boundary} in this networks
is the emptyset. Hence, boundary need to be built. The building process chosen
consists in adding structurally specific nodes~\cite{Martinelli1994}. This leads
to the following definitions, considering an arbitrary TBANs $N$ whose
underlying digraph $G=(V,A)$ is such that $V \subset \mathbb{Z}^2$ and that $V^c
= \mathbb{Z}^2 \setminus V$ is the set of vertices of $N^c$, said to be the
complement of $N$ in $\mathbb{Z}^2$.

\begin{definition}
  \label{def_externalboundary}
  The \emph{external boundary} (called \emph{boundary} for short), denoted by
  $\partial_{\mathrm{ext}}N$, is the set of nodes of $N^c$ at distance $1$ (in
  terms of distance in $\mathbb{Z}^2$) to at least one node of $N$ such that:
  \begin{equation*}
    \partial_{\mathrm{ext}}N = \{i \in N^c \; | \; \exists j \in N: i \in {\cal
      N}_j, j \notin {\cal N}_i\}\text{.}
  \end{equation*}
\end{definition}

\begin{figure}
  \centerline{
    \includegraphics[scale=1]{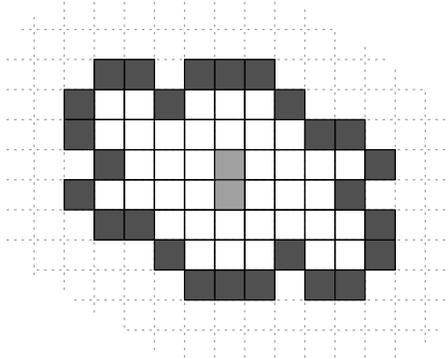}
  }
  \caption{An arbitrary TBAN $N$ on $\mathbb{Z}^2$ whose nodes are in white and
    light grey (in the case of central nodes) and boundary
    $\partial_{\mathrm{ext}}N$ is the set of nodes coloured in dark grey.}
  \label{fig_boundary}
\end{figure}

An illustration of centre and boundary of a TBAN on $\mathbb{Z}^2$ is given in
Figure~\ref{fig_boundary}.\medskip

TBANs in the sequel are \emph{attractive}, \textit{i.e.}, they are such that
$w_{i,i} < 0$ and $\forall j \in \Lambda_i, j\neq i, w_{i,j} > 0$. Note also
that activation thresholds are all fixed to $0$ and that auto-interaction
potentials are always taken into account. Thus, the $w_{i,i}$'s play the role of
activation thresholds. Furthermore, as said in the introduction, nonlinearity is
added in the model of TBANs considering that interaction potentials that act on
a node $i$ at time $t$ are not only reduced to the combination of the
auto-interaction potential $w_{i,i}$ and the nearest-neighbours potential
$\sum_{j\in \Lambda_i} w_{i,j} \cdot x_j(t)$. Indeed, we consider also
\emph{coalition potentials}. For instance, given a node $i$ of a TBAN $N$ at
time $t$ whose state is not known, if we consider that the evolution of node $i$
takes into account coalition of neighbours couples, the interaction potential of
node $i$ equals $w_{i,i} + \sum_{j\in \Lambda_i} w_{i,j} \cdot x_j(t) +
\sum_{j,\ell \in {\cal N}_i} w_{i,\langle j,\ell \rangle} \cdot x_j(t) \cdot
x_{\ell}(t)$, where $w_{i,\langle j,\ell \rangle}$ defines the interaction
weight that the couple of active nodes $j$ and $\ell$ has on $i$. Remark that
the $w_{i,i}$'s correspond to thresholds (considering them separately) and that
the $i$'s play the role of elements of coalitions (considering them as parts of
couples, triples, quadruples and quintuples in the sequel) \medskip

Let $T \in \mathbb{R^+}$ be the \emph{temperature parameter}. We give the
following notations of interaction potentials for every node $i$ of an arbitrary
TBAN $N$ to ease the reading:
\begin{itemize}
\item $u_{0,i} = \frac{w_{i,i}}{T}$, called \emph{singleton potential},
  a function of the auto-interaction weight of an arbitrary node $i$ (always
  taken into account);
\item $u_{1,i,j} = \frac{w_{i,j}}{T}$, where $j \, \in \, \Lambda_i$,
  \emph{couple potential}, a function of interaction weights received by node
  $i$ from its strict nearest neighbours;
\item $u_{2,i,\langle j,\ell\rangle} = \frac{w_{i,\langle j,\ell\rangle}}{T}$,
  where $j,\ell \, \in \, {\cal N}_i \; ; \; j\neq \ell$, called \emph{triple
    potential}, a function of interaction weights received by node $i$ from
  couples of its active neighbours;
\item $u_{3,i,\langle j,\ell,m\rangle} = \frac{w_{i,\langle
      j,\ell,m\rangle}}{T}$, where $j,\ell,m \, \in \, {\cal N}_i \; ; \; j\neq
  \ell \neq m$, called \emph{quadruple potential}, a function of interaction
  weights received by node $i$ from triples of its active neighbours;
\item $u_{4,i,\langle j,\ell,m,p\rangle} = \frac{w_{i,\langle
      j,\ell,m,p\rangle}}{T}$, where $j,\ell,m,p \, \in \, {\cal N}_i \; ; \;
  j\neq \ell\neq m\neq p$, called \emph{quintuple potential}, a function of
  interaction weights received by node $i$ from quadruples of its active
  neighbours.
\end{itemize}

\begin{definition}
  \label{def_NTBAN}
  A \emph{stochastic TBAN $N$ of order $k$} on $\mathbb{Z}^2$ is a TBAN whose
  local transition function $f_i$ calculates the probability for node $i$ to be
  at state $1$ at time $t+1$ knowing the configuration projected on its
  neighbourhood ${\cal N}_i$ at time $t$ and taking into account $1$-uple,
  $2$-uple, \ldots, $k$-uple potentials, with $2 \leq k \leq 5$ such that:
  \begin{equation}
    \label{eq_NN2}
    \forall i \in N = \{1, \ldots, n\}, P(x_i(t+1)=\alpha) = \frac{e^{\alpha
        \cdot (u_{0,i} + \sum_{j \in \Lambda_i} u_{1,i,j} \cdot x_j(t) +
        \phi_i^k(\Lambda_i))}} {1 + e^{u_{0,i} + \sum_{j \in \Lambda_i}
        u_{1,i,j} \cdot x_j(t) + \phi_i^k(\Lambda_i)}}\mathrm{,}
  \end{equation}
  where $\phi_i^k(\Lambda_i)$ is the nonlinear term such that:
  \begin{multline*}
    \phi_i^k(\Lambda_i) = \sum_{\stackrel{\stackrel{j,\ell,m,p,q \in {\cal
            N}_i}{j \neq \ell \neq m \neq p \neq q}}{}} u_{2,i,\langle j,\ell
      \rangle} \cdot x_j(t) \cdot x_{\ell}(t) + \ldots +\\
    u_{k-1,i,\langle j,\ell,m,p,q\rangle} \cdot x_j(t) \cdot x_{\ell}(t) \cdot
    x_m(t) \cdot x_p(t) \cdot x_q(t)\mathrm{.}
  \end{multline*}
\end{definition}

Remark that, in the case of TBANs of order $2$ (\textit{i.e.}
$\phi_i^k(\Lambda_i)=0$), if $T$ tends to $0$, then the stochastic local
transitions functions defined in~\ref{eq_NN2} are equivalent to the
deterministic one defined in Equation~\ref{eq_NN1}. Before going further, let us
insist that, from Definition~\ref{def_NTBAN}, we derive that nonlinear TBANs
studied in this report are stochastic TBANs of order at least equal to $3$.

\section{Theoretical approach and phase transitions}
\label{res}

Let us recall that TBANs considered in the sequel are isotropic, translation
invariant, nonlinear. Moreover, we add that they are attractive. Given a
stochastic TBAN $N$, that means that $\forall i \in N, \, u_{0,i} < 0 \; ; \;
\forall j \in \Lambda_i, \, u_{1,i,j} > 0$.

\subsection{Projectivity matrix}

\begin{definition}
  A \emph{cylinder} $[A,B]$ is a configuration x such that:
  \begin{equation*}
    [A,B] = \{x \; | \; \forall i \in A, \, x_i = 1; \; \forall i\in B, \, 
    x_i = 0\}\text{.}
  \end{equation*}
\end{definition}

If $\mu$ denotes the invariant measure of a stochastic TBAN $N$ composed of $n$
nodes, indexed from $1$ to $n$, such that $n$ tends to infinity, we have the
following projectivity and conditional relations. Indeed, we can write
\emph{projectivity equations} such that:
\begin{multline*}
  \forall A, B \subset N \; | \; A \cap B = \emptyset, \, \forall i \in A,\\
  \mu([A,B]) + \mu([A \setminus \{i\}, B \cup \{i\}]) = \mu([A \setminus 
  \{i\}, B])\text{,}
\end{multline*}
where $\mu([A,B])$ is the probability to observe the configuration $[A,B]$. We
calso write \emph{conditional equations} (\textit{i.e.}, the Bayes formulas)
such that:
\begin{equation}
  \label{eq_cond}
  \forall i \in N,\, \mu([\{i\}, \emptyset]) = \sum_{A,B \subset 
    N \; | \; A \cap B = \emptyset, \, A \cup B = N \setminus \{i\}} 
  \Phi_i(A,B) \cdot \mu([A,B])\text{,}
\end{equation}
where $\Phi_i(A,B)$ denotes the conditional probability that state of node $i$
equals $1$ knowing cylinder $[A, B]$ such that:
\begin{equation*}
  \mu(x_i = 1 \; | \; [A,B]) = \Phi_i(A,B) = 
  \frac{e^{u_{0,i} + \sum_{j \in \Lambda_i} u_{1,i,j} \cdot x_j(t) + \phi_i^k(\Lambda_i))}}
  {1 + e^{u_{0,i} + \sum_{j \in \Lambda_i} u_{1,i,j} \cdot x_j(t) + 
      \phi_i^k(\Lambda_i)}}\text{.}
\end{equation*}

Consider $L = (N \cup \partial_{\text{ext}}N) \setminus \{O\}$ such that nodes
of $L$ are ordered according to the lexical order of their indices. For every
subset $K$ of $L$ of size $k$, we denote by $j_K$ the minimal index of nodes
belonging to $K$. \emph{Projectivity matrix} $M$ of order $2^{|L|}$ is defined
such that \textit{(i)} the $2^{|L|}-1$ first lines contain respectively the
coefficients of the projectivity equations for any of the $2^{|L|}-1$ different
couple $[L,K]$ and \textit{(ii)} the last line contains the coefficients of the
conditional equation $\mu([\{O\}, \emptyset]) = \sum_{A,B \subset N~|~A \cap B =
  \emptyset, A \cup B = N \setminus \{O\}} \Phi(A,B) \cdot \mu([A,B])$ that
calculates the global probability for the central node to be active. The system
of equations obtained from the projectivity and conditional equations is:
\begin{equation}
  \label{eq_syst}
  M \cdot \hspace*{-3pt}\begin{pmatrix}
    \mu([L,\emptyset]) \\
    \mu([L \setminus \{1\}, \{1\}]) \\
    \mu([L \setminus \{2\}, \{2\}]) \\
    \vdots \\
    \mu([K, L \setminus K]) \\
    \mu([K \setminus \{j_K\}, (L \setminus K) \cup \{j_K\}]) \\
    \vdots \\
    \mu([\{1\}, L \setminus \{1\}]) \\
    \mu([\emptyset, L]) \\
  \end{pmatrix}
  =
  \begin{pmatrix}
    \mu([L \setminus \{1\}, \emptyset]) \\
    \mu([L \setminus \{2\}, \emptyset]) \\
    \ldots \\
    \vdots \\
    \mu([K \setminus \{j_K\}, L \setminus K]) \\
    \ldots \\
    \vdots \\
    \mu([\emptyset, L \setminus \{1\}]) \\
    \mu([\{O\}, \emptyset]) \\
  \end{pmatrix}\text{.}
\end{equation}

From this system of equations, it is easy to write: 
\begin{equation*}
  M = \begin{pmatrix}
    1&1&0&0&0&0&\ldots&0&0&0\\
    1&0&1&0&0&0&\ldots&0&0&0\\
    1&0&0&1&0&0&\ldots&0&0&0\\
    1&0&0&0&1&0&\ldots&0&0&0\\
    0&1&0&0&0&1&\ldots&0&0&0\\
    \vdots&\vdots&\vdots&\vdots&\vdots&\vdots&\ddots&
    \vdots&\vdots&\vdots\\
    0&0&0&0&0&0&\ldots&0&1&0\\
    0&0&0&0&0&0&\ldots&0&0&1\\
    \Phi_0&\Phi_1&\Phi_2&\Phi_3&\Phi_4&\Phi_5&\ldots&\Phi_{13}&\Phi_{14}&\Phi_{15}
  \end{pmatrix}\text{,}
\end{equation*}
where $\Phi_0 = \Phi(\Lambda_O, \emptyset)$, $\Phi_1 = \Phi(\Lambda_O \setminus
\{1\}, \{1\})$, $\Phi_2 = \Phi(\Lambda_O \setminus \{2\}, \{2\})$, \ldots,
$\Phi_5 = \Phi(\Lambda_O \setminus \{1,2\}, \{1,2\})$, \ldots, $\Phi_{13} =
\Phi(\Lambda_O \setminus \{1,3,4\}, \{1,3,4\})$, \ldots and $\Phi_{15} =
\Phi(\emptyset, \Lambda_O)$..\medskip

Projectivity and conditional equations are in general linearly
independent. However, under specific parametric conditions such as conditions of
non uniqueness of the invariant measure, that is not the case. From the work of
Dobrushin in~\cite{Dobrushin1968a,Dobrushin1968b,Dobrushin1968c,Dobrushin1969}
in the framework of random fields, we derive the following definition.

\begin{definition}
  \label{def_phtr}
  Let $N$ be an arbitrary stochastic attractive TBAN. Let
  $\partial_{\mathrm{ext}}^0N$ (resp. $\partial_{\mathrm{ext}}^1N$) be a
  boundary of $N$ composed of nodes whose state is fixed to $0$ (resp. $1$). The
  dynamical behaviour of $N$ admits a \emph{phase transition} if and only if the
  invariant measure of the Markov chain associated to $N
  \cup \partial_{\mathrm{ext}}^0N$ does not equals that of the Markov chain
  associated to $N \cup \partial_{\mathrm{ext}}^1N$.
\end{definition}

From Equations~\ref{eq_syst} and Definition~\ref{def_phtr}, we can directly
write the following proposition.

\begin{proposition}
  \label{prop_detM}
  Given $N$ a stochastic attractive TBAN, the nullity of the determinant of its
  associated projectivity matrix $M$ is a necessary condition for $N$ to admit a
  phase transition in its dynamical behaviour.
\end{proposition}

\begin{lemma}
  \label{lemma_determinant}
  \cite{demongeot1981}~The nullity of the determinant of a projectivity matrix
  is characterised by:
  \begin{equation*}
    \text{\emph{Det}}M = 0 \iff \sum_{K \subset L} (-1)^{|L \setminus K|} \cdot 
    \Phi(K, L \setminus K) = 0\text{.}
  \end{equation*}
\end{lemma}

Because of our hypotheses of isotropy and translation invariance, it is
interesting to note that we can use the spatial Markovian property in order to
make easier solving the system of projectivity equations. The \emph{spatial
  Markovian property} implies that the state of the centre $O$ of a network $N$
depends only on the states of its neighbours, which allows to reduce $L$ to the
centre $O$ strict neighbourhood, namely $\Lambda_O = {\cal N}_O \setminus
\{O\}$. Then, it is simpler to build the associated projectivity matrix $M_O$ of
order $2^{2 \cdot d}$.

\subsection{Results}

Basing our approach on Proposition~\ref{prop_detM}, in this section, we prove
the existence of parametric conditions of stochastic nonlinear TBANs that admit
phase transitions.\medskip

First, from the \emph{spatial Markovian property} of TBANs and because
$|\Lambda_O| = 0 \mod 2$, the right member of the equation of
Lemma~\ref{lemma_determinant} can be written pairing the subsets $K$ and
$\Lambda_O \setminus K$, namely considering that:
\begin{multline*}
  (-1)^{|\Lambda_O \setminus K|} \cdot \Phi(K, \Lambda_O \setminus K) +
  (-1)^{|K|} \cdot \Phi(\Lambda_O \setminus K, K)\\
  = (-1)^{|K|} [\Phi(K, \Lambda_O \setminus K) + \Phi(\Lambda_O \setminus K,
  K)]\text{.}
\end{multline*}

By hypothesis, nonlinear term $\phi_O^k(K)$ is symmetric and equals $-2\cdot
u_{0,O} - \sum_{j\in \Lambda_O} u_{1,O,j} - \phi_O^k(\Lambda_O \setminus
K)$. The symmetry property of the nonlinear term means that $\phi_O^k(K) =
\phi_O^k(\Lambda_O) - \phi_O^k(\Lambda_O \setminus K)$.

\begin{lemma}
  \label{lemma_sym}
  Given $N$ a nonlinear TBAN of order $k$ and $\phi_O^k(K) = -2\cdot u_{0,O} -
  \sum_{j\in \Lambda_O} u_{1,O,j} - \phi_O^k(\Lambda_O \setminus K)$ a symmetric
  nonlinear term such that $\phi_O^k(K) = \phi_O^k(\Lambda_O) -
  \phi_O^k(\Lambda_O \setminus K)$, we have:
  \begin{multline}
    \label{eq_sym}
    \phi_O^k(K) = \phi_O^k(\Lambda_O) - \phi_O^k(\Lambda_O \setminus K) \iff\\
    u_{0,O} + \frac{\sum_{j\in \Lambda_O} u_{1,O,j}}{2} +
    \frac{\phi_O^k(\Lambda_O)}{2} = 0\text{.}
  \end{multline}
\end{lemma}

\begin{proof}
  Let us note $\phi_O^k(\Lambda_O) - \phi_O^k(\Lambda_O \setminus K) =
  \phi_{\text{sym}}$. Trivially, developing the left member of
  Equation~\ref{eq_sym} by definition of nonlinear terms, we can write:
  \begin{equation*}
    \begin{split}
      \phi_O^k(K) = \phi_{\text{sym}}\iff\:& -2\cdot u_{0,O}-
      \sum_{j\in\Lambda_O}u_{1,O,j} -\phi_O^k(\Lambda_O \setminus K) =
      \phi_{\text{sym}}\\
      \iff\:& -2\cdot u_{0,O}-\sum_{j\in \Lambda_O}u_{1,O,j} =\phi_O^k(\Lambda_O)\\
      \iff\:& -2\cdot u_{0,O}-\sum_{j\in \Lambda_O}u_{1,O,j} -\phi_O^k(\Lambda_O)=0\\
      \iff\:& -u_{0,O}-\frac{\sum_{j\in \Lambda_O}u_{1,O,j}}{2} 
      -\frac{\phi_O^k(\Lambda_O)}{2}=0\\
      \iff\:& u_{0,O}+\frac{\sum_{j\in\Lambda_O}u_{1,O,j}}{2}
      +\frac{\phi_O^k(\Lambda_O)}{2}=0\text{,}
    \end{split}
  \end{equation*}
  which is the expected result.
\qquad\end{proof}

\begin{lemma}
  \label{lemma_proba}
  Let $N$ be a nonlinear TBAN of order $k$ and $\phi_O^k(\Lambda_O) = -2\cdot
  u_{0,O} - \sum_{j\in \Lambda_O} u_{1,O,j}$ be the nonlinear term of $N$ when
  every nearest neighbour of its central node $O$ is active. Then:
  \begin{equation*}
    u_{0,O} + \sum_{j \in \Lambda_O} \frac{u_{1,O,j}}{2} + 
    \frac{\phi_O^k(\Lambda_O)}{2} = 0 \iff \Phi(K, \Lambda_O \setminus K) + 
    \Phi(\Lambda_O \setminus K, K) = 1\text{.}
  \end{equation*}
\end{lemma}
\begin{proof} First, let us show that $\Phi(K, \Lambda_O \setminus K) +
  \Phi(\Lambda_O \setminus K, K) = 1$. It suffices to multiply $\Phi(K,
  \Lambda_O \setminus K)$ by $1 = \frac{e^{-2\cdot u_{0,O} - \sum_{j \in
        \Lambda_O} u_{1,O,j} - \phi_O^k(\Lambda_O)}}{e^{-2\cdot u_{0,O} -
      \sum_{j \in \Lambda_O} u_{1,O,j} - \phi_O^k(\Lambda_O)}}$:
  \begin{equation*}
    \Phi(K, \Lambda_O \setminus K) = \frac{e^{u_{0,O} + \sum_{j \in K}
        u_{1,O,j} + \phi_O^k(K)}} {1+e^{u_{0,O} + \sum_{j \in K} u_{1,O,j} +
        \phi_O^k(K)}}
    \times \frac{e^{-2\cdot u_{0,O} - \sum_{j \in \Lambda_O}
        u_{1,O,j} - \phi_O^k(\Lambda_O)}}{e^{-2\cdot u_{0,O} - \sum_{j \in
          \Lambda_O} u_{1,O,j} - \phi_O^k(\Lambda_O)}}\text{.}
  \end{equation*}
  Given $\delta$ defined by:
  \begin{equation*}
    \delta = e^{-2\cdot u_{0,O} - \sum_{j \in \Lambda_O} u_{1,O,j} -
      \phi_O^k(\Lambda_O)} + e^{-u_{0,O} - \sum_{j \in \Lambda_O \setminus K}
      u_{1,O,j} + \phi_O^k(K)- \phi_O^k(\Lambda_O)}\text{,}
  \end{equation*}
  we have:
  \begin{equation*}
    \Phi(K, \Lambda_O \setminus K) = \frac{e^{-u_{0,O} - \sum_{j \in
          \Lambda_O \setminus K} u_{1,O,j} + \phi_O^k(K) -
        \phi_O^k(\Lambda_O)}}{\delta}\text{.}
  \end{equation*}
  By hypothesis, $\phi_O^k(\Lambda_O) = -2\cdot u_{0,O} - \sum_{j \in \Lambda_O}
  u_{1,O,j}$. As a consequence, we have $e^{-2\cdot u_{0,O} - \sum_{j \in \Lambda_O}
    u_{1,O,j} - \phi_O^k(\Lambda)} = 1$. Moreover, given that nonlinear term
  $\phi_O^k$ is symmetric:
  \begin{equation*}
    \begin{split}
      \Phi(K, \Lambda_O \setminus K) & = \frac{e^{-u_{0,O} - \sum_{j \in
            \Lambda_O \setminus K} u_{1,O,j} - \phi_O^k(\Lambda_O \setminus
          K)}}{1+e^{-u_{0,O} - \sum_{j \in \Lambda_O \setminus K} u_{1,O,j} -
          \phi_O^k(\Lambda_O \setminus K)}}\\
      & = 1 - \frac{e^{u_{0,O} + \sum_{j \in \Lambda_O \setminus K} u_{1,O,j} +
          \phi_O^k(\Lambda_O \setminus K)}}{1 + e^{u_{0,O} + \sum_{j \in
            \Lambda_O \setminus K} u_{1,O,j} +
          \phi_O^k(\Lambda_O \setminus K)}}\\
      & = 1 - \Phi(\Lambda_O \setminus K, K)\text{.}
    \end{split}
  \end{equation*}
  So, we can write:
  \begin{equation*}
    \Phi(K, \Lambda_O \setminus K) + \Phi(\Lambda_O \setminus K, K) = 1 \iff
    \Phi(\Lambda_O \setminus K, K) = 1 - \Phi(K, \Lambda_O \setminus 
    K)\text{.}
  \end{equation*}
  Expanding left and right members of the equation above leads to:
  \begin{equation*}
    \frac{e^{u_{0,O} + \sum_{j \in \Lambda_O \setminus K} u_{1,O,j} +
        \phi_O^k(\Lambda_O \setminus K)}}{1 + e^{u_{0,O} + \sum_{j \in
          \Lambda_O \setminus K} u_{1,O,j} + \phi_O^k(\Lambda_O \setminus K)}}
    = 
    1 - \frac{e^{u_{0,O} + \sum_{j \in K} u_{1,O,j} + \phi_O^k(K)}}{1 +
      e^{u_{0,O} + \sum_{j \in K} u_{1,O,j} + \phi_O^k(K)}}\text{,}
  \end{equation*}
  which is equivalent to:
  \begin{equation*}
    \frac{e^{u_{0,O} + \sum_{j \in \Lambda_O \setminus K} u_{1,O,j} +
        \phi_O^k(\Lambda_O \setminus K)}}{1 + e^{u_{0,O} + \sum_{j \in
          \Lambda_O \setminus K} u_{1,O,j} +
        \phi_O^k(\Lambda_O \setminus K)}} = 
    \frac{e^{-u_{0,O} - \sum_{j \in K} u_{1,O,j} - \phi_O^k(K)}}{1 + e^{-u_{0,O}
        - \sum_{j \in K} u_{1,O,j} - \phi_O^k(K)}}\text{.}
  \end{equation*}
  Let us proceed to the following change of variables: let $\delta_1$
  (resp. $\delta_2$) be the denominator of the left member (resp. of the right
  member) and $\eta_1$ (resp. $\eta_2$) the numerator of the left member
  (resp. of the right member) of the equation above. We have then:
  \begin{equation*}
    \begin{split}
      \frac{\eta_1}{\delta_1} = \frac{\eta_2}{\delta_2} & \iff \frac{\eta_1\cdot
        \delta_2}{\delta_1\cdot\delta_2} = \frac{\eta_2\cdot\delta_1}
      {\delta_2\cdot\delta_1}\\
      & \iff \eta_1\cdot \delta_2 = \eta_2\cdot \delta_1\text{.}
    \end{split}
  \end{equation*}
  Let $\zeta$ be such that:
  \begin{equation*}
    \zeta = e^{\sum_{j \in \Lambda_O \setminus K}
      u_{1,O,j} - \sum_{j \in K} u_{1,O,j} + \phi_O^k(\Lambda_O \setminus K) -
      \phi_O^k(K)}\text{.}
  \end{equation*}
  We have:
  \begin{equation*}
    \begin{split}
      \frac{\eta_1}{\delta_1} = \frac{\eta_2}{\delta_2} & \iff \eta_1 + \zeta =
      \eta_2 + \zeta\\
      & \iff \eta_1 = \eta_2\text{.}
    \end{split}
  \end{equation*}
  Thus, we can write:
  \begin{multline*}
    \frac{\eta_1}{\delta_1} = \frac{\eta_2}{\delta_2} \iff e^{u_{0,O} +
      \sum_{j \in \Lambda_O \setminus K} u_{1,O,j} + \phi_O^k(\Lambda_O
      \setminus K)} = e^{-u_{0,O} - \sum_{j \in K} u_{1,O,j} - \phi_O^k(K)}\\
    \iff  u_{0,O} + \sum_{j \in \Lambda_O \setminus K} u_{1,O,j} +
    \phi_O^k(\Lambda_O \setminus K) = -u_{0,O} - \sum_{j \in K} u_{1,O,j} -
    \phi_O^k(K)\\
    \iff \phi_O^k(K) = -2\cdot u_{0,O} - \sum_{j \in \Lambda_O \setminus K}
    u_{1,O,j} -\sum_{j \in K}u_{1,O,j}-\phi_O^k(\Lambda_O \setminus K)\text{.}
  \end{multline*}
  And, thus, we have:
  \begin{equation*}
    \frac{\eta_1}{\delta_1} = \frac{\eta_2}{\delta_2} \iff \phi_O^k(K) = -2\cdot
    u_{0,O} - \sum_{j \in \Lambda_O} u_{1,O,j} - \phi_O^k(\Lambda_O \setminus
    K)\text{.}
  \end{equation*}
  Hence, by hypothesis:
  \begin{equation*}
    \begin{split}
      \frac{\eta_1}{\delta_1} = \frac{\eta_2}{\delta_2} & \iff \phi_O^k(K) =
      \phi_O^k(\Lambda_O) - \phi_O^k(\Lambda_O \setminus K)\text{,}\\
    \end{split}
  \end{equation*}
  which is the expected result.
\qquad\end{proof}

From Lemmas~\ref{lemma_sym} and~\ref{lemma_proba}, it is easy to derive the
following theorem that highlights an empirical sufficient condition of phase
transitions in nonlinear TBANs of order $k$ on $\mathbb{Z}^d$.

\begin{theorem}
  \label{thm_det} Let $N$ be a nonlinear TBAN of order $k$. We have:
  \begin{equation*}
    \phi_O^k(K) = \phi_O^k(\Lambda_O) - \phi_O^k(\Lambda_O \setminus K) \implies
    \text{Det}M = 0\text{,}
  \end{equation*}
  which means that the symmetry property of the non linear term is an empirical
  sufficient condition for detM to vanish, allowing consequently phase
  transitions to occur.
\end{theorem}

\begin{proof}
  From Lemma~\ref{lemma_determinant} and because of the parity of the cardinal
  of $\Lambda_O$, we can write:
  \begin{equation*}
    \begin{split}
      \text{\emph{Det}}M = 0 \iff\: & \sum_{K \subset \Lambda_O}
      (-1)^{|\Lambda_O
        \setminus K|} \cdot \Phi(K, L \setminus K) = 0\\
      \iff\: & \sum_{K \subset \Lambda_O} (-1)^{|\Lambda_O \setminus K|} \times
      \frac{[\Phi(K, L \setminus K)+\Phi(L \setminus K, K)]}{2} = 0\text{.}
    \end{split}
  \end{equation*}
  Then Lemma~\ref{lemma_proba} leads to:
    \begin{equation*}
    \begin{split}
      \text{\emph{Det}}M = 0 & \iff \sum_{K \subset \Lambda_O} (-1)^{|\Lambda_O
        \setminus K|} \cdot \frac{1}{2} = 0\text{,}\\
    \end{split}
  \end{equation*}
  which is always true. As a result, since Lemmas~\ref{lemma_sym}
  and~\ref{lemma_proba} are based on the hypothesis of symmetry of the non
  linear term, we have from Lemma~\ref{lemma_determinant}:
  \begin{equation*}
    u_{0,O} + \frac{\sum_{j\in \Lambda_O} u_{1,O,j}}{2} +
    \frac{\phi_O^k(\Lambda_O)}{2} = 0 \implies \text{\emph{Det}}M = 0\text{,}
  \end{equation*}
  which is the expected result.
\qquad\end{proof}

\bibliography{ds2010}

\end{document}